\documentstyle[twoside,fleqn,espcrc2,epsf]{article}

\newcommand{\Eta}{H}
\newcommand{\assign}{=}

\newcommand{\trace}{{\rm Tr}}

\newcommand{\A}{M}

\newcommand{\Src}{\Eta}
\newcommand{\src}{\eta}
\newcommand{\Sln}{\Psi}
\newcommand{\sln}{\psi}

\newcommand{\nn}{\nonumber}
\newcommand{\nest}{\hspace*{5mm}}
\newcommand{\keysolid}{---}
\newcommand{\keydots}{$\cdot \cdot \cdot$}
\newcommand{\keydotdashdot}{$\cdot - \cdot$}
\newcommand{\keyshortdash}{- - -}
\newcommand{\keylongdash}{-- --}

\newcommand{\beqn}{\begin{equation}}
\newcommand{\eeqn}{\end{equation}}
\newcommand{\eqref}[1]{Eq.~(\ref{#1})}

\newcommand{\AmS}{{\protect\the\textfont2
  A\kern-.1667em\lower.5ex\hbox{M}\kern-.125emS}}

\title{Block Algorithms for Quark Propagator Calculation}
\author{Stephen M. Pickles\address{Department of Physics and Astronomy, 
University of Edinburgh,
Edinburgh EH9 3JZ, Scotland}
, UKQCD Collaboration}

\begin{document}


\begin{abstract}
Computing quark propagators in lattice QCD is equivalent to solving 
large, sparse linear systems with multiple right-hand sides. 
Block algorithms attempt to accelerate the convergence of iterative
Krylov-subspace methods by solving the multiple systems simultaneously.
This paper compares a block generalisation of the quasi-minimal residual
method (QMR), Block Conjugate Gradient on the normal equation, 
Block Lanczos and ($\gamma_5$-symmetric) Block BiConjugate Gradient.
\end{abstract}

\maketitle

\vspace{-1ex}
\section{Introduction}
\vspace{-1ex}
The calculation of quark propagators remains a major bottleneck in lattice
QCD. The problem is to solve the matrix equation
\begin{equation} \label{TheEquations}
        \A \sln = \src
\end{equation}
for $\sln$
given several right-hand sides ({\em sources}) $\src$.
The fermion matrix $\A$ is non-hermitian 
(we use Wilson fermions with clover improvement), 
sparse and very large.
Recent attempts to accelerate the solution of \eqref{TheEquations} 
have focused on:\\
\nest 1. improved iterative methods\\
\nest 2. improved preconditioners, or\\
\nest 3. solving related systems.

There is growing consensus that the first of these areas is now mature
\cite{Forcrand,FrommerReview}, with BiCGSTAB 
emerging as the method to beat. 

Finding a better preconditioner is complicated by the now universal
requirement of scalability on parallel computers; 
red-black preconditioning (used in the present study) is beginning to give 
way to LL-SSOR \cite{Fischer}, but the last word on preconditioning
has not been said.

The third area is the exploitation of information
found in the solution of one system to accelerate the convergence of another.
Keeping the source fixed and varying $\kappa$ leads
to a family of multiple mass tricks \cite{ManyMasses,Glassner,Beat}. 
Keeping the matrix fixed and varying the source leads to the ideas of deflation
\cite{Forcrand} and block algorithms, the several systems being solved
simultaneously in the latter but sequentially in the former.

Two properties of $\A$ are relevant here.
\vspace{-1ex}
\begin{enumerate} \setlength{\parskip}{0 ex}
\item \label{SymmetryProp}
$\A = \gamma_5 \A^\dagger \gamma_5 $ is $\gamma_5$-symmetric.
This has been exploited to halve the computational costs
of QMR and BiCG \cite{ManyMasses}.
\item \label{ShiftedProp}
$\frac{1}{\kappa} \A$
is a shifted matrix with respect to its dependence on $\frac{1}{\kappa}$.
Multiple-mass tricks depend on this property.
\end{enumerate}
\vspace{-1ex}
Red-black preconditioning preserves property (\ref{SymmetryProp}),
but destroys property (\ref{ShiftedProp}) except in the unimproved case
of $C_{SW}=0$.

\vspace{-1ex}
\section{Block Algorithms}
\vspace{-1ex}
Using a Krylov subspace method $s$ times to solve $s$ systems
$\A \sln^{(1)} = \src^{(1)}, \ldots, \A \sln^{(s)} = \src^{(s)}$
leads to the construction of several overlapping Krylov subspaces.
In the worst case, ie. when the number of iterations required for
convergence equals the order $N$ of $\A$, the overlap will be complete.
By solving the $s$ systems simultaneously, block algorithms eliminate
the redundant matrix-vector operations in the above approach.
One assembles the $s$ right-hand sides into an $N \times s$ matrix
$ \Src = ( \src^{(1)}, \ldots, \src^{(s)} ) $
and solves
\beqn \label{BlockEquation}
\A \Sln = \Src 
\eeqn 
for
$ \Sln = ( \sln^{(1)}, \ldots, \sln^{(s)} ) $.

On the other hand,
a perfect preconditioner (one which coincides exactly with $\A^{-1}$),
solves the system with a single multiplication; 
in this case there is no gain to be had from the block algorithm.
In practice we hope to have good preconditioners, so that we usually
solve the point ($s=1$) problem to the desired accuracy in much less 
than $N$ multiplications. 
These considerations lead one to expect that blocking will be
most effective on badly conditioned systems and/or small volumes, ie.
when the number of iterations required for the point algorithm
to converge is comparable to $N$.

Blocking introduces certain overheads. The first is memory; 
the storage requirements for vectors in the point algorithm
are multiplied by $s$ when going to the block algorithm. 
Secondly, vector-vector operations such as
\beqn
y = \alpha x + y \ {\rm and} \ \beta = y^\dagger x \nn
\eeqn
in the point algorithm generalise to 
\beqn
Y = X \alpha + Y \ {\rm and} \ \beta = Y^\dagger X \nn
\eeqn
where $X$ and $Y$ are $N \times s$ matrices and $\alpha$ and $\beta$
have become $s \times s$ matrices. Thus the number of
vector-vector operations required per iteration scales as $s^2$.

\vspace{-1ex}
\subsection{B-CGNR and B-Lanczos}
Block Conjugate Gradient \cite{OLeary} can be applied to the normal equation
$\A^\dagger \A \Sln = \A^\dagger \Src$.
Increasing $s$ yields a clear improvement in convergence, but not enough to
defray the cost of squaring the condition number. 
Reference \cite{Henty2} studied a method based on the hermitian block Lanczos
process and applied it to 
$\gamma_5 \A \Sln = \gamma_5 \Src$.
B-Lanczos clearly outperforms B-CGNR.
Unfortunately, $\gamma_5$ is a bad
preconditioner for \eqref{BlockEquation}.

\vspace{-1ex}
\subsection{Block B-BiCG($\gamma_5$)}
The algorithm presented here is a special case of, and easily derived from, 
the Block Bi-Conjugate Gradient algorithm of O'Leary \cite{OLeary}.
I have used $\gamma_5$-symmetry to eliminate multiplications by 
$\A^\dagger$, and have followed her important suggestion of 
orthonormalising the columns of $P_k$.

\begin{center}
\vspace{3pt} {\bf B-BiCG($\gamma_5$)} \\
\end{center}
\begin{eqnarray}
&& R_0 \assign \Src - \A \Sln_0 \nn \\
&& \rho_0 \assign R_0^\dagger \gamma_5 R_0 \nn \\
&& P_0 \delta_0 \assign R_0 \label{bicg_g5_qr_a} \\
&& {\rm for\ } k = 0, 1, 2, \ldots 
{\rm \ until\ convergence\ do\ } \{ \nn \\
&& \nest T \assign \A P_k \nn \\
&& \nest \alpha_k \assign \left( P_k^\dagger \gamma_5 T \right)^{-1} 
	\delta_k^{-\dagger} \rho_k \nn \\
&& \nest \Sln_{k+1} \assign P_k \alpha_k + \Sln_k \nn \\
&& \nest R_{k+1} \assign -T \alpha_k + R_k \nn \\
&& \nest \rho_{k+1} \assign R_{k+1}^\dagger \gamma_5 R_{k+1} \nn \\
&& \nest \beta_k \assign \delta_k \rho_k^{-1} \rho_{k+1} \nn \\
&& \nest T \assign R_{k+1} + P_k \beta_k \nn \\
&& \nest P_{k+1} \delta_{k+1} \assign T \label{bicg_g5_qr_b} \\
&& \} \nn
\end{eqnarray}
The operations (\ref{bicg_g5_qr_a}) and (\ref{bicg_g5_qr_b}) 
are $QR$ decompositions, as are 
(\ref{BQMRMGSa}), (\ref{BQMRMGSb}) and  (\ref{BQMRQR}) below. 

\vspace{-1ex}
\subsection{Block QMR}

The original QMR (Quasi-Minimal Residual) algorithm
\cite{FreundAndNachtigal} used blocks of variable sizes
in look-ahead steps to avoid Lanczos breakdown.  Boyse and Seidl \cite{Boyse} 
described a block version of QMR for complex symmetric matrices,
using fixed-size blocks to accelerate convergence.
Subsequently Freund and Malhotra \cite{FreundAndMalhotra}
discovered a non-hermitian version.
The version presented here was developed by me independently 
of \cite{FreundAndMalhotra}.

\begin{center}
\vspace{3pt} {\bf B-QMR($\gamma_5$)}
\end{center}
\begin{eqnarray}
&& P_0            \assign P_{-1} \assign V_0 \assign 0 \nn \\
&& c_0            \assign b_{-1} \assign b_0 \assign 0;
\  a_0            \assign d_{-1} \assign d_0 \assign I \nn \\
&& R_0            \assign \tilde{V}_1 \assign \Src - \A \Sln_0 \nn \\
&& V_1 \rho_1     \assign \tilde{V}_1 \label{BQMRMGSa} \\
&& \tilde{\tau}_1 \assign \rho_1 \nn \\
&& {\rm for\ } k = 1, 2, \ldots 
{\rm \ until\ convergence\ do\ } \{ \nn \\
&& \nest \delta_k           \assign V_k^\dagger \gamma_5 V_k \nn \\
&& \nest \beta_k            \assign \delta_{k-1}^{-1} \rho_k^\dagger \delta_k \nn \\
&& \nest T                  \assign \A V_k - V_{k-1} \beta_k \nn \\
&& \nest \alpha_k           \assign \delta_k^{-1} V_k^\dagger \gamma_5 T \nn \\
&& \nest \tilde{V}_{k+1}    \assign T - V_k \alpha_k \nn \\
&& \nest V_{k+1} \rho_{k+1} \assign \tilde{V}_{k+1} \label{BQMRMGSb} \\
&& \nest \theta_k        \assign b_{k-2} \beta_k \nn \\
&& \nest \epsilon_k      \assign a_{k-1} d_{k-2} \beta_k + b_{k-1} \alpha_k \nn \\
&& \nest \tilde{\zeta}_k \assign c_{k-1} d_{k-2} \beta_k + d_{k-1} \alpha_k \nn \\
&& \nest
\left( \begin{array}{cc} a_k & b_k \\ c_k & d_k \end{array} \right) ^\dagger
\left( \begin{array}{c} \zeta_k \\ 0 \end{array} \right)
\assign
\left( \begin{array}{c} \tilde{\zeta}_k \\ \rho_{k+1} \end{array} \right) 
\label{BQMRQR} \\
&& \nest P_k                \assign (V_k - P_{k-1} \epsilon_k 
                             - P_{k-2} \theta_k) \zeta_k^{-1} \nn \\
&& \nest \tau_k             \assign a_k \tilde{\tau}_k;
\ \tilde{\tau}_{k+1} \assign c_k \tilde{\tau}_k \nn \\
&& \nest \Sln_k             \assign \Sln_{k-1} + P_k \tau_k \nn \\
&& \} \nn
\end{eqnarray}

The algorithm does not give a recurrence for the residual $R_k$,
but the fact that $\trace\ \tilde{\tau}_k^\dagger \tilde{\tau}_k$
is of the same order of magnitude as $\trace\ R_k^\dagger R_k$
is useful in formulating a stopping criterion.

This version does nothing to address the problems of Lanczos breakdown
and (near) linear dependence in the columns of $V_k$. 
These can be brought under control, as shown in \cite{FreundAndMalhotra}.

\begin{minipage}{72mm}
\begin{center}
\leavevmode
{\setlength{\epsfxsize}{70mm} \setlength{\epsfysize}{60mm}
\epsfbox{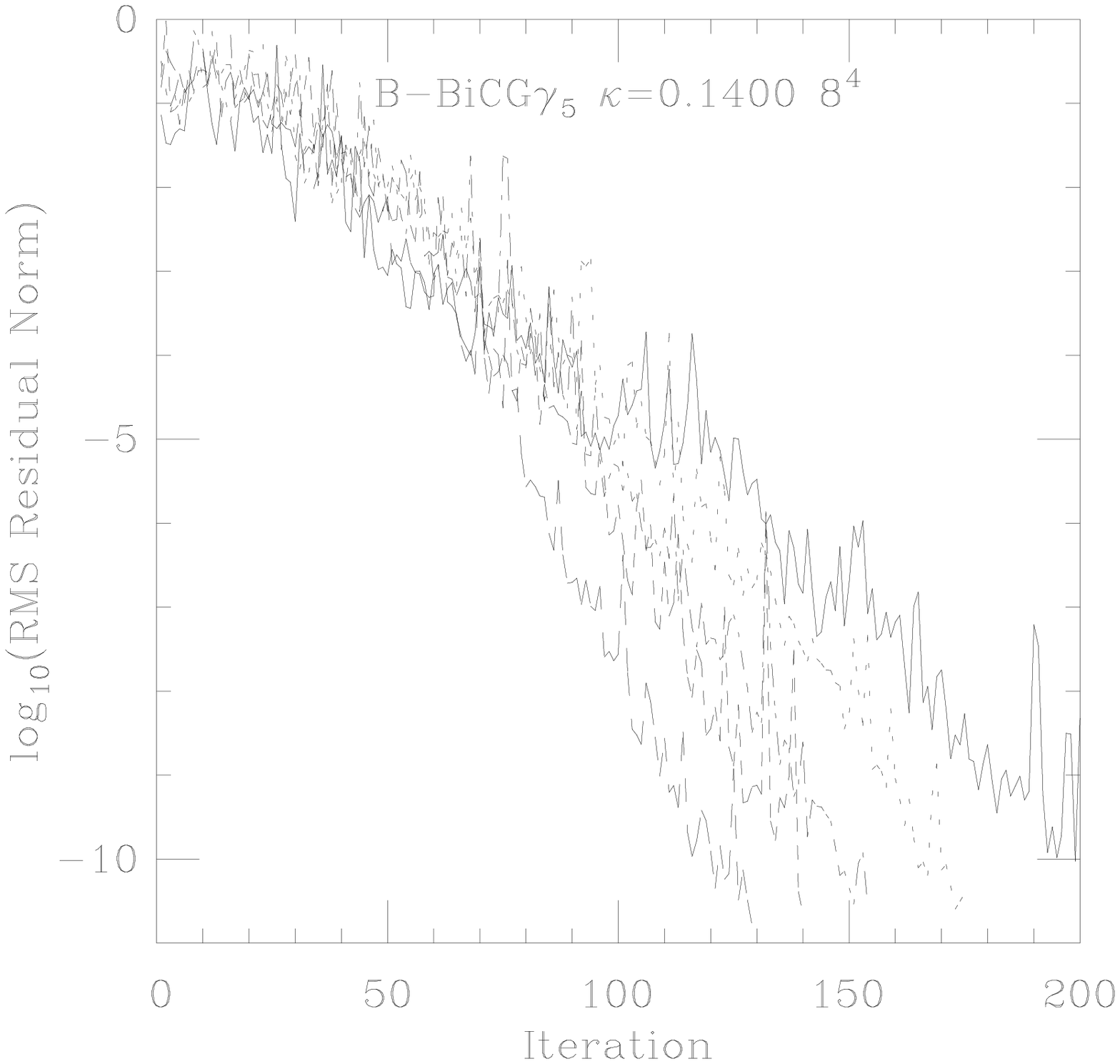}}
{\setlength{\epsfxsize}{70mm} \setlength{\epsfysize}{60mm}
\epsfbox{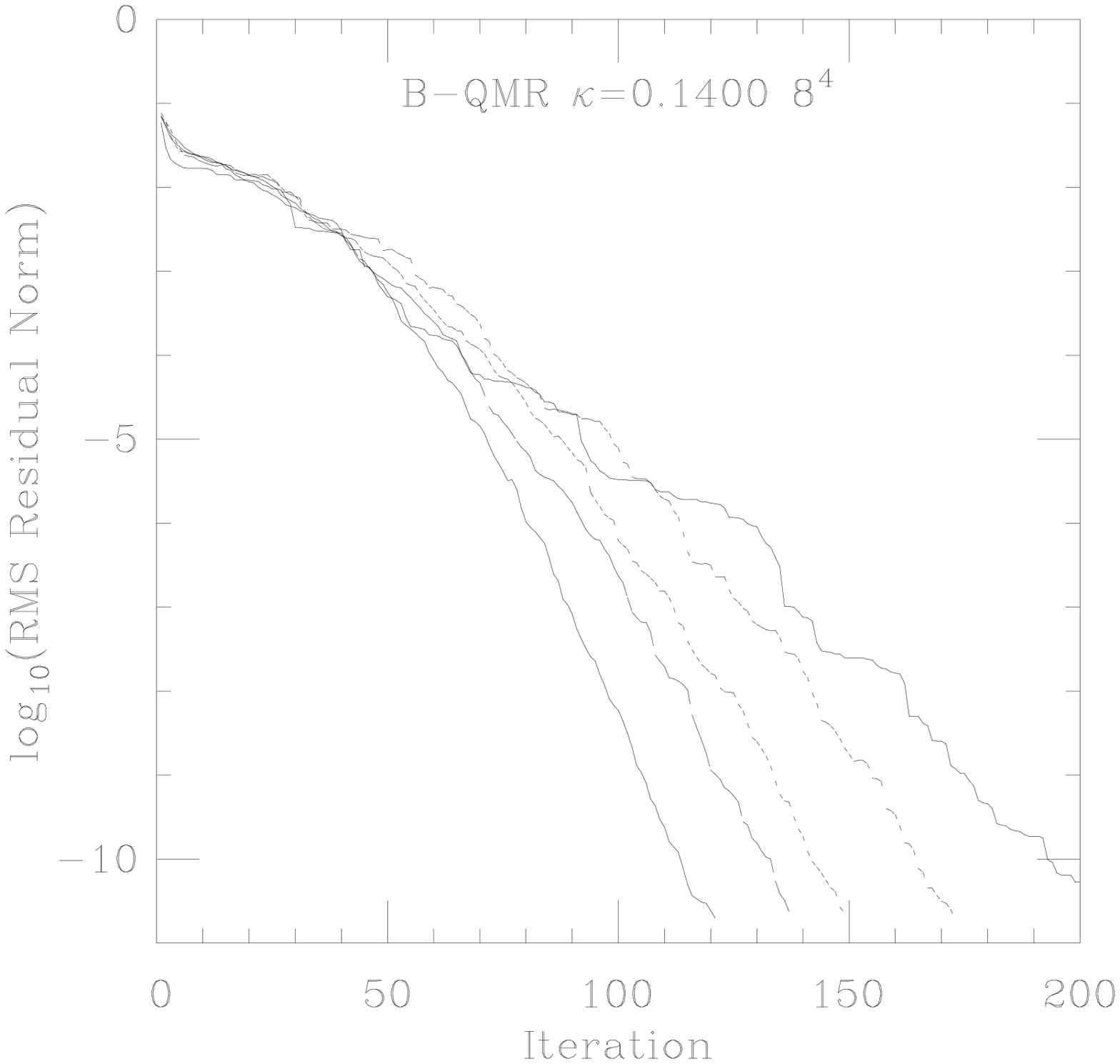}}
{\small
Convergence histories for B-BiCG$\gamma_5$, B-QMR$\gamma_5$.\\
$s = 1$ (\keysolid), $2$ (\keydots), $3$ (\keydotdashdot),
$4$ (\keyshortdash), $6$ (\keylongdash); \\
$\kappa = 0.14$, 
$V=8^4$, $\beta=5.7$,
$C_{SW} = 1.5678$.
}
\end{center}
\end{minipage}

\vspace{-1ex}
\section{Conclusions}
\vspace{-1ex}
Block algorithms can reduce the number of matrix-vector operations
required for convergence at the expense of more vector-vector operations.
This does not necessarily lead to a reduction in wall-clock time. 

B-QMR($\gamma_5$) and B-BiCG($\gamma_5$) are clearly faster than 
B-CGNR and B-Lanczos, 
and in some regimes (light masses and small volumes) they
significantly outperform BiCGSTAB.

However, on lattices of realistic size, the improvements from blocking
are marginal at best.
At $\beta=6.0$ quenched, $V=16^3 \times 48$, I found that 
$s=1$ is near-optimal for both 
B-QMR($\gamma_5$) and B-BiCG($\gamma_5$), unless the configuration
is exceptional. 
At $\beta=5.2$, $N_f=2$, $V=12^3 \times 24$, 
I found $s=1$ to be optimal for the same algorithms
at all $\kappa_{valence}$ and $\kappa_{sea}$ combinations studied in
\cite{Talevi}.
These conclusions should be reviewed if the relative cost of matrix-vector
to vector-vector operations increases significantly. Block algorithms
may yet have a role to play in conjunction with highly-improved actions
on coarse lattices.

\vspace{-1ex}
\section*{ACKNOWLEDGMENTS}
\vspace{-1ex}
I thank R.~Freund, U.~Gl\"{a}ssner and D.~Henty for helpful communications and
discussions. I used resources funded by 
PPARC grant GR/L22744 and EPSRC grants GR/K41663 and GR/K55745.



\end{document}